\documentclass[aps,prl,twocolumn,nopacs,floatfix]{revtex4-2}
\usepackage{epsfig}
\usepackage{graphicx}
\usepackage{dcolumn}
\usepackage{amsthm,amsmath}
\usepackage{comment}
\usepackage[colorlinks=true, allcolors=blue]{hyperref}
\usepackage{xcolor}
\usepackage{mathtools}
\usepackage{orcidlink}
\usepackage{amsmath}
\usepackage[normalem]{ulem}
\usepackage{bm}

\begin{document}

\title{Precise Determination of Electric Quadrupole Moments and Isotope Shift Constants of Yb$^+$ in Pursuance of Probing Fundamental Physics and Nuclear Radii} 

\author{B. K. Sahoo}
\email{bijaya@prl.res.in}
\affiliation{Atomic and Molecular Physics Division, Physical Research Laboratory, Navrangpura, Ahmedabad 380009, India}

\date{\today}

\begin{abstract}
Contemplating to register signature of a new vector boson unambiguously from the measured non-linear isotope shift (IS) effects in three recent experiments [Phys. Rev. X {\bf 12}, 021033 (2022); Phys. Rev. Lett. {\bf 128}, 163201 (2022) and 
Phys. Rev. Lett. {\bf 134}, 063002 (2025)], very precise values of quadrupole moments and IS constants for the $6s ~ ^2S_{1/2} \rightarrow 5d ~ ^2D_{3/2}$ and $6s ~ ^2S_{1/2} \rightarrow 5d ~ ^2D_{5/2}$ clock transitions of $^{171}$Yb$^+$ are presented. This is accomplished by incorporating contributions from the computationally challenging triply excited configurations through the relativistic coupled-cluster (RCC) theory. Testament of quality atomic wave functions of states of the above transitions, obtained using the RCC theory, are gauged by comparing the calculated energies and magnetic dipole hyperfine structure constants with their measurements. The improved quadrupole moments from this work will be immensely useful to estimate quadrupole shifts of the clock transitions of Yb$^+$. Complementary approaches are employed to ascertain accuracy and comprehend roles of orbital relaxation and correlation effects in evaluating the IS constants. Combining these constants with the IS measurements from Phys. Rev. Lett. {\bf 128}, 163201 (2022), differential nuclear charge radii of the Yb isotopes are inferred that deviate by 6-7\% from the literature data.
\end{abstract}

\maketitle

The singly charged ytterbium (Yb$^+$) is one of the most favourable ions by the experimentalists. Long lifetime of its $[4f^{13}6s^2] ^2F_{7/2}$ state (1.58(8) year) \cite{Lange} offers a very narrow transition to the ground state for carrying out clock frequency measurement to high-precision \cite{Tofful, tamm}. Similarly, lifetimes of the next two metastable states $[4f^{14}5d] ^2D_{3/2}$ and $[4f^{14}5d] ^2D_{5/2}$ of Yb$^+$ are about 55 $ms$ and 7 $ms$, respectively \cite{yu, schacht}. Thus, the $4f^{14}6s~^2S_{1/2} \rightarrow 4f^{14}5d~^2D_{3/2}$ (denoted as $\alpha$) and $4f^{14}6s~^2S_{1/2} \rightarrow 4f^{14}5d~^2D_{5/2}$ (denoted by $\beta$) transitions, decay to the ground state mainly via the electric quadrupole (E2) channel, are also considered to be suitable for atomic clocks \cite{Huntemann-PRA-2014, Lange1}. Simultaneous clock frequency measurements in these transitions from the same ion could significantly reduce systematic uncertainties. Since Yb$^+$ is a heavier atomic system with its valence electron in the $s$ orbital, it is obvious to exhibit strong relativistic effects. These characteristics make Yb$^+$ a special clock candidate to be used for probing temporal and spatial variation of the fine-structure constant \cite{Lange1}. The other notable applications using the above transitions are for making qubits with longer coherence-time \cite{Wang}, testing parity nonconservation \cite{bijaya1, rahaman}, probing local Lorentz symmetry violation \cite{dzuba} etc. to name a few.

Recently, Yb$^+$ is on spotlight for inferring signature of existence of a new vector boson from its non-linear effects in the isotope shift (IS) measurements \cite{Ono,Counts,hur,Door}. {\it Albeit} measurements are performed to high-precision, atomic factors considered in the analyses differ significantly from various methods \cite{Counts, hur, Door, Allehabi}. The calculations mostly aimed at deducing field shift (FS) constants, which are relatively easier to compute than the normal mass shift (NMS) and specific mass shift (SMS) constants of IS in Yb$^+$ \cite{bijaya2}. It is, therefore, the need of hour to calculate the IS constants of this ion more reliably. Also, precise estimate of systematic effects in the IS measurements are equally important in order to probe new physics. One of the major systematic effects in these measurements stem from electric quadrupole (E2) shifts in the $5d~^2D_{3/2}$ and $5d~^2D_{5/2}$ states \cite{Counts,hur,Door}. Their precise values are greatly useful in the clock frequency measurements \cite{bijaya3,itano}. E2 shifts can be estimated with the knowledge of quadrupole moments ($\Theta$s) of atomic states. There is lack of experimental value of $\Theta$ for the $5d~^2D_{5/2}$ state, while two experiments offer contrasting $\Theta$ values of the $5d~^2D_{3/2}$ state \cite{Schneider, Lange2}. In this case too, substantial differences among the calculations using distinctive methods are noticed \cite{bijaya3, Guo, ymyu}. 

In this Letter, we intend to produce improved values of the IS constants and $\Theta$ of the states of both the $\alpha$ and $\beta$ transitions in Yb$^+$. This is attained by embodying the computationally challenging triple excitations in the calculations using the relativistic coupled-cluster (RCC) theory. To corroborate the claimed accuracy of our calculations, we also determine electron attachment (EA) energies and magnetic dipole hyperfine structure constants ($A_{hf}$) and compare them with their experimental results. Reliabilities of the IS constant calculations are further ascertained by evaluating them using two complementary approaches in the RCC theory. In fact, it is imperative to determine the $F$ constants of Yb$^+$ assuredly for precise estimations of nuclear charge radii of the Yb isotopes.   

\begin{table}[t]
\centering
\caption{Calculated energy (EA), $A_{hf}$ and $\Theta$ values at different levels of approximation and comparison of final values with precise experiments (Expt).} 
\begin{tabular}{l ccc } \hline \hline
 Method &  $6s~^2S_{1/2}$  & $5d~^2D_{3/2}$ & $5d~^2D_{5/2}$  \\ \hline 
 \multicolumn{4}{c}{Energies (in cm$^{-1}$)} \\
 DHF   &  $-90790$ & $-66517$ & $-66038$ \\
 RCCSD &  $-97642$ & $-73525$ & $-72268$ \\
 RCCSDT & $-97859$ & $-74198$ & $-72915$ \\
 $+$Extra & $-191$ & $-752$ & $-703$ \\
 $+$Breit & 68 & $-2$ & $-28$ \\
  & & & \\
  Final  & $-97982(300)$ & $-74952(350)$ & $-73646(350)$ \\
  Expt \cite{nist} & $-98232$  & $-75271$ & $-73899$ \\
  \hline 
     \multicolumn{4}{c}{$A_{hf}$ constants (in MHz)} \\
 DHF &  9753.18 & 290.74 & 110.64 \\
 RCCSD & 13096.52 & 400.20 & $-66.78$ \\
 RCCSDT & 13157.79 & 423.60 & $-72.32$ \\
 $+$Extra & 53.91 & 1.48 & 3.84 \\
 $+$Breit & $-29.77$ & 2.09 & 0.78 \\
 $+$BW &  $-316.62$ & $-2.16$ & 3.40 \\
  & & & \\
  Final  & 12865(200) & 425(7) & $-64(3)$ \\
  Expt \cite{martensson,engelke,roberts} & 12645 & 430(43) & $-63.6(5)$ \\
    \hline 
   \multicolumn{4}{c}{$\Theta$ value (in a.u.)} \\
 DHF &   & 2.500 & 3.690 \\
 RCCSD &  & 2.076 & 3.125 \\
 RCCSDT &  & 2.027 & 3.053 \\
 $+$Extra &  & $-0.051$ & $-0.074$ \\
 $+$Breit &  & 0.001 & $-0.002$ \\
  & & & \\
  Final  &  & 1.977(11) & 2.976(15) \\
  Ref. \cite{ymyu} &  & 1.973(31) &  3.06(7)  \\
  Expt \cite{Schneider, Lange2} &  & 2.08(11), 1.95(1) & \\
\hline \hline
\end{tabular}
\label{tab1}
\end{table}

Due to unique adeptness to capture electron correlation effects and capable of elucidating roles of different physical effects in lucid manner, either through hole-particle excitation processes or using diagrammatic techniques \cite{bartlett}, the RCC theory is preferred over other accessible methods to calculate properties in many electron systems. Besides its size-consistent characteristic, variant of tailor-made procedures can be adopted in the RCC theory to determine a wide range of properties deftly. In the RCC theory {\it ans\"atz}, wave function ($|\Psi_0 \rangle$) of the ground state is expressed by (e.g. see \cite{bartlett})
\begin{eqnarray}
|\Psi_0 \rangle = e^S |\Phi_0 \rangle ,
\end{eqnarray}  
where ($|\Phi_0 \rangle$ is the mean-field (MF) wave function and $S$ generates excitation determinants from $|\Phi_0 \rangle$. The Dirac-Hartree-Fock (DHF) method is used to obtain the MF wave function with atomic Hamiltonian ($H$) incorporating the Dirac-Coulomb as well as the Breit interactions in the self-consistent manner. The orbitals from the $s$, $p_{1/2;3/2}$, $d_{3/2;5/2}$, $f_{5/2;7/2}$, $g_{7/2;9/2}$, $h_{9/2;11/2}$ and $i_{11/2;13/2}$ symmetries were constructed by taking linear combinations of 40 Gaussian type orbitals for each spinor and their radial components were extended till 500 atomic units (a.u.) by respecting the kinetic balance condition.

\begin{table}[t]
\centering
\caption{Calculated FS, NMS and SMS constants from the FF approach at different levels of approximation.}
\begin{tabular}{l ccc } \hline \hline
 Method &  $6s~^2S_{1/2}$  & $5d~^2D_{3/2}$ & $5d~^2D_{5/2}$  \\ \hline 
 \multicolumn{4}{c}{FS constants (in MHz fm$^{-2}$)} \\
 DHF   &  $-12378.38$ & 1800.63 & 1453.22  \\
 RCCSD &  $-14639.81$ & 1720.27 & 1458.33  \\
 RCCSDT & $-14701.06$ & 1568.51 & 1363.24 \\
 $+$Extra & $-76.56$  & 31.83 & 26.61 \\
 $+$Breit & 84.22 & $-17.33$ & 6.86 \\
  & & & \\
  Final  & $-14693.40$ & 1593.01 & 1396.71 \\
  \hline 
   \multicolumn{4}{c}{NMS constants (in GHz amu)} \\
 DHF &  1457.34 &  1093.84 & 1091.00   \\
 RCCSD &  1562.01 & 1202.50 & 1191.62 \\
 RCCSDT & 1558.79 & 1208.07 & 1197.32\\
 $+$Extra & 3.43 & 11.28 & 11.04 \\
 $+$Breit & $-2.83$ & 0.44 & 1.78 \\
  & & & \\
  Final  & 1559.39 & 1219.79 & 1210.14 \\
  \hline 
     \multicolumn{4}{c}{SMS constants (in GHz amu)} \\
 DHF &  $-1037.40$ & $-1254.62$ & $-1309.41$  \\
 RCCSD &  $-100.86$ & $-653.15$ & $-787.93$\\
 RCCSDT & $-65.84$ & $-622.56$ & $-726.67$ \\
 $+$Extra & $-5.54$ & $-23.59$ & $-24.75$ \\
 $+$Breit & 2.65 & $-11.91$ & $-10.84$ \\
  & & & \\
  Final  & $-68.73$ & $-658.06$ & $-762.26$\\
\hline \hline
\end{tabular}
\label{tab2}
\end{table}

\begin{table}[t]
\centering
\caption{Calculated FS, NMS and SMS constants from the AR approach at different levels of approximation.}
\begin{tabular}{l ccc } \hline \hline
 Method &  $6s~^2S_{1/2}$  & $5d~^2D_{3/2}$ & $5d~^2D_{5/2}$  \\ \hline 
 \multicolumn{4}{c}{FS constants (in MHz fm$^{-2}$)} \\
 DHF   &  $-11327.28$ & $\sim 0.0$ & $\sim 0.0$ \\
 RCCSD &  $-14662.60$ & 1712.92 & 1492.97 \\
 RCCSDT &  $-14739.94$  & 1435.22 & 1230.69 \\
 $+$Extra & $-68.16$  & 37.47 & 32.38 \\
 $+$Breit & 50.12 & 8.90 & 12.36 \\
  & & & \\
  Final  & $-14657.98$ & 1481.59 & 1275.43 \\
  \hline 
   \multicolumn{4}{c}{NMS constants (in GHz amu)} \\
 DHF &   3755.62 &  5847.74 &  5524.06   \\
 RCCSD &  1415.46 &  1200.35 & 1201.46 \\
 RCCSDT & 1505.51 & 1188.77 & 1188.61 \\
 $+$Extra & 9.31 & 19.88 & 18.01  \\
 $+$Breit & 0.84 &  3.98 & 4.47 \\
  & & & \\
  Final  & 1515.66 & 1212.63 & 1211.09 \\
  \hline 
     \multicolumn{4}{c}{SMS constants (in GHz amu)} \\
 DHF &  $-2455.69$ & $-4106.98$ & $-3996.33$    \\
 RCCSD & $-16.16$  & $-756.36$ & $-899.44$ \\
 RCCSDT & 65.87 & $-499.66$ & $-624.95$ \\
 $+$Extra & $-10.28$ & $-30.37$ & $-30.04$ \\
 $+$Breit & 3.19 & $-13.35$ & $-11.61$\\
  & & & \\
  Final  & 58.78 & $-543.38$ & $-666.60$ \\
\hline \hline
\end{tabular}
\label{tab3}
\end{table}

Following the Fock-space depiction of the RCC theory \cite{bijaya2,bijaya3,lindgren}, wave functions of the $6s ~ ^2S_{1/2}$ and $5d ~ ^2D_{3/2;5/2}$ states with the common closed-core, $[4f^{14}]$, are expressed as
\begin{eqnarray}
|\Psi_v \rangle = e^S |\Phi_v  \rangle = e^T \{1+S_v\} |\Phi_v  \rangle,
\end{eqnarray}
where $|\Phi_v  \rangle= a_v^{\dagger} |\Phi_0 \rangle$ and $S$ is divided into $T$ that takes into account correlations from the core and $S_v$ to take care of the correlations including the valence orbitals $v=6s$ and $v=5d_{3/2;5/2}$. For computational efficacy, $T$ amplitudes of the common core are obtained first and subsequently they are used to obtain the $S_v$ amplitudes (details are given elsewhere; e.g. see \cite{bijaya2,bijaya3}). With $|\Phi_0 \rangle$ as the Fermi vacuum, normal order formalism is adopted to estimate the EA ($E_v$) directly. Calculations at the singles and doubles approximation in the RCC theory (RCCSD method) and singles, doubles and triples approximation of the RCC theory (RCCSDT method) are performed by allowing orbitals up to $g$-symmetry. Contributions from the $h$- and $i$-symmetry orbitals are given at the RCCSD level separately under $+$Extra. This facilitates to circumvent the computational limitation to include triple excitations and highlights importance of using orbitals from the higher angular momentum symmetry to produce accurate results in Yb$^+$. Also, additional corrections arising from the Breit ($+$Breit) interactions are estimated at the RCCSD method. 

The $A_{hf}$ and $\Theta$ values are estimated using the expectation value evaluation (EVE) expression 
\begin{eqnarray}
 \langle O \rangle \equiv \frac{\langle \Psi_v | O | \Psi_v \rangle}{\langle \Psi_v | \Psi_v \rangle} = \frac{\langle \Phi_v | \{1+S_v \}^{\dagger}  \bar{O} \{ 1+ S_v \} |\Phi_v \rangle} {\langle \Phi_v | \{1+ S_v \}^{\dagger} \bar{N} \{ 1+ S_v \} |\Phi_v \rangle}, \ \ \
 \label{eqeve}
\end{eqnarray}
where $O$ denotes the operator of the respective property and contributions from $\bar{O}=e^{T^{\dagger}}O e^T$ and $\bar{N}=e^{T^{\dagger}} e^T$ were accounted through an iterative procedure \cite{bijaya3}.

\begin{table}[t]
\centering
\caption{Comparison of differential IS constants of the $\alpha$ and $\beta$ transitions from different theoretical methods and approaches with experiments.}
\begin{tabular}{l cc } \hline \hline
 Quantity &  Value & Method, Approach \\ \hline 
 $F_{\alpha}$ (in MHz fm$^{-2}$)  &  $-16286(50)$ & RCC, FF (this work)  \\
                   & $-16139(200)$  & RCC, AR (this work) \\
                    & $-14910.0$  & CI, FF \cite{Door} \\
                    & $-13070.0$  & CI, EVE \cite{Door} \\
                    & $-18720.1$  & RPA$+$BO, FF \cite{Allehabi} \\
                    & $-16094.0$ & CI, FF \cite{Counts, hur} \\
                    & $-16771.0$ & MBPT, FF \cite{Counts} \\
                    & $-14968.0$ & CI$+$MBPT, FF \cite{hur} \\
 $F_{\beta}$ (in MHz fm$^{-2}$)  & $-16090(50)$  & RCC, FF (this work) \\ 
                 &  $-15933(200)$ & RCC, AR (this work) \\
                 & $-14690.0$ & CI, FF \cite{Door} \\
                 & $-13090.0$ & CI, EVE \cite{Door} \\
                 & $-18302.6$ & RPA$+$BO, FF \cite{Allehabi} \\
                 & $-15852.0$ & CI, FF \cite{Counts} \\
                 & $-16570.0$ & MBPT, FF \cite{Counts} \\
                 & $-14715.0$ & CI$+$MBPT, FF \cite{hur} \\
 $F_{\alpha \beta}$  & 1.0122 & RCC, FF (this work) \\
                     & 1.0129  & RCC, AR (this work) \\
                     & 1.0153 &  CI, FF \cite{Counts, hur} \\
                     & 1.0121 &  MBPT, FF \cite{Counts} \\
                     & 1.0172 & CI$+$MBPT, FF \cite{hur} \\
                     & 1.01141024(86) & Experiment \cite{Counts, hur} \\
 $K_{\alpha}^{\rm{NMS}}$ (in GHz amu) & 340(50) & RCC, FF (this work) \\
                     &  303(70)  & RCC, AR (this work) \\
                      &  377.61 & Scaling \cite{nist} \\
 $K_{\beta}^{\rm{NMS}}$ (in GHz amu) & 349(50) & RCC, FF (this work) \\
                     & 305(70) & RCC, AR (this work) \\
                     &   400.17 & Scaling \cite{nist} \\  
 $K_{\alpha}^{\rm{SMS}}$ (in GHz amu) & 589(30) & RCC, FF (this work) \\
                                      & 602(50) &  RCC, AR (this work) \\
 $K_{\beta}^{\rm{SMS}}$ (in GHz amu) & 693(30) & RCC, FF (this work) \\
                                     & 725(50)   &  RCC, AR (this work) \\
 $K_{\alpha}^{\rm{MS}}$ (in GHz amu) & 929(60) & RCC, FF (this work) \\
                              &  905(86) &  RCC, AR (this work) \\
                     & $1638.5$ & CI, FF \cite{Counts, hur} \\
                     & $661$ & CI$+$MBPT, FF \cite{hur} \\
 $K_{\beta}^{\rm{MS}}$ (in GHz amu) & 1042(60) & RCC, FF (this work)\\
                               & 1030(86)  &  RCC, AR (this work) \\
                      & $1678.2$ & CI, FF \cite{Counts, hur} \\
                      & $752$ & CI$+$MBPT, FF \cite{hur} \\
$K_{\alpha \beta}^{MS}$  & $-124.89$ & RCC, FF (this work) \\
                         & $-136.75$  &  RCC, AR (this work) \\
                      & $-65$  &  CI, FF \cite{Counts, hur} \\
                      & $-103.92$ & CI$+$MBPT, FF \cite{hur} \\
                      & $-120.208(23)$ & Experiment \cite{Counts} \\
\hline \hline
\end{tabular}
\label{tab4}
\end{table}

The $E_v$, $A_{hf}$ and $\Theta$ values along with  uncertainties to the final results for the ground and $5D$ states of Yb$^+$ are presented at the DHF, RCCSD and RCCSDT methods in Table \ref{tab1}. The uncertainties are stemmed from the basis functions as well as from the neglected contributions. The $+$Extra and $+$Breit contributions are also listed in the same table. $+$Extra contributions to EAs suggest correlation effects arising through the high-lying orbitals are more crucial than the higher-order relativistic effects, which are often neglected due to the computational hardship involved in accounting them in a heavier system like Yb$^+$. Our final EA values are in good agreement with the experimental results tabulated in the National Institute of Science and Technology (NIST) database \cite{nist}. From the $A_{hf}$ values, we find both the electron correlation effects and higher-order relativistic effects are equally responsible for their accurate evaluations. The Bohr-Weisskopf effects to $A_{hf}$ (given as $+$BW) are also observed to be quite significant and can be said to be the reason for large discrepancies seen between the earlier calculations and experimental values \cite{martensson,engelke,roberts,bijaya3}. In contrast, $+$Breit corrections are found to be insignificant to $\Theta$ of the $5d~^2D_{3/2}$ and $5d~^2D_{5/2}$ states as seen from Table \ref{tab1}. It, thus, means that accuracy of the calculated $\Theta$ values entirely depend on how well electron correlation effects are treated in the method. This is also evident from the differences in the RCCSD and RCCSDT values and $+$Extra contributions to the $\Theta$ values. In fact, $+$Extra contributions are very prominent. Our final values seem to be in agreement with the experimental values \cite{Schneider, Lange2} as well as with our previous estimation using the finite-field (FF) approach \cite{ymyu} for the $5D$ states. In the FF approach, it yields
\begin{eqnarray}
\langle O \rangle = \left. \frac{\partial E_v(\lambda)}{\partial \lambda} \right|_{\lambda=0}  \approx \frac{ E_v(+\lambda)-E_v(-\lambda)}{2 \lambda} ,
\label{eqn1}
\end{eqnarray}
where $E_v(\lambda)$ is the EA due to $H(\lambda)= H + \lambda O$ for a small parameter $\lambda$. In Ref. \cite{ymyu}, orbitals were spanned over radial range $\le 50$ a.u. against 500 a.u. in the present work, triple excitations were neglected for the $5d ~ 5D_{5/2}$ state and they were added using a smaller basis function in the $5d ~ 5D_{3/2}$ state. Results from other theoretical studies are discussed in Refs. \cite{bijaya3,ymyu} and they seem to overestimate the values.

Discussions on various results from Table \ref{tab1} affirm that the calculated atomic wave functions from the RCCSDT method are able to provide credible results for properties sensitive to both inside and in the vicinity of the nuclear region as well as in the asymptotic region. The auxiliary contributions from $+$Extra and $+$Breit refine the results further. In the interest of comprehending accuracy of the calculated IS constants, it would be intriguing to compare results from different many-body methods as well as from different approaches in the RCC theory. This is owing to the fact that electron correlations are channelized differently through these approaches and methods. For instance, estimating properties in the FF approach take care of orbital relaxation effects, deals with finite number of terms and satisfy the Hellmann-Feynman theorem \cite{bartlett} but this approach can contain errors \cite{bijaya2,bijaya00} due to use of an arbitrary parameter for numerical differentiation in Eq. (\ref{eqn1}). On other hand, EVE suffers from brute-force termination of non-terminating expressions, not following the Hellmann-Feynman theorem and necessitates to include orbital relaxation effects through random phase approximation (RPA) like terms \cite{bijaya44} but it is free from any numerical differentiation errors. These shortcomings of EVE are addressed by our recently developed analytical response (AR) approach in the RCC theory \cite{bijaya4,bijaya5}, in which $\langle O \rangle$ is evaluated by 
\begin{eqnarray}
(H-E_v) |\Psi_v^{(1)} \rangle = (\langle O \rangle - O ) |\Psi_v \rangle ,
\end{eqnarray}
where $|\Psi_v^{(1)} \rangle$ is the first-order perturbed wave function due to $O$ and can be used to estimate the second-order corrections to probe non-linear IS effects.

\begin{table}[t]
\centering
\caption{The inferred differential nuclear charge radii $\delta \langle r^2 \rangle$ between isotopes $A$ and $A'$ ranging $168-176$ of Yb from the IS measurements in the $\alpha$ and $\beta$ transitions (indicated with subscripts $\alpha$ and $\beta$) of Yb$^+$ \cite{Counts}. The averaged value from both the transitions is recommended ($\delta \langle r^2 \rangle_{\rm{reco}}$) as the final value and compared with the literature (Lit.) data.}
\begin{tabular}{c cccc } \hline \hline
 ($A$, $A'$) &  $\delta \langle r^2 \rangle_{\alpha} $ & $\delta \langle r^2 \rangle_{\beta} $  & $\delta \langle r^2 \rangle_{\rm{reco}}$ & Lit. \cite{Angeli} \\ \hline 
(168, 170) & $-0.1463$ & $-0.1463$ & $-0.1463(7)$ & $-0.1561(3)$ \\
(170, 172) & $-0.1375$ & $-0.1375$ & $-0.1375(7)$ & $-0.1479(1)$ \\
(172, 174) & $-0.1074$ & $-0.1074$ & $-0.1074(6)$ & $-0.1207(1)$ \\
(174, 176) & $-0.1025$ & $-0.1025$ & $-0.01025(6)$ & $-0.1159(1)$ \\
\hline \hline
\end{tabular}
\label{tab5}
\end{table}  

Using a combined RPA and Br\"uckner orbitals (RPA$+$BO) method, the FS constants ($F$) for the $\alpha$ and $\beta$ transitions are obtained as $-18720.1$ and $-18302.6$ MHz/fm$^2$, respectively, with spherical nucleus approximation \cite{Allehabi}. In Ref. \cite{Counts}, these values are renewed to $-1609.4$ and $-1677.1$ MHz/fm$^2$ using the configuration interaction (CI) method, and $-1585.2$ and $-1657.0$ MHz/fm$^2$ using the many-body perturbation theory (MBPT) method. Similarly, almost 40\% differences between the differential mass shift (MS) constants, NMS and SMS constants together, are seen between the experimental and calculations \cite{Counts, hur}. We determine these constants using both the FF and AR approaches in the RCC theory. The values from the FF approach are presented in Table \ref{tab2} considering $\lambda=1.0 \times 10^{-5}$ a.u., while results from the AR approach are given along with their uncertainties in Table \ref{tab3} at different levels of approximation. We find consistent among these values. Differential IS constants of the $\alpha$ and $\beta$ transitions ($O_{i}$ denotes $\langle O \rangle$ constant for $i$-transition) from this work and earlier studies \cite{Counts, hur, Allehabi, Door} are given in Table \ref{tab4}. Since $F_{\alpha \beta} = F_{\alpha}/F_{\beta}$ and $K^{MS}_{\alpha \beta} = K^{MS}_{\alpha} - F_{\alpha \beta} K^{MS}_{\beta}$ are also measured, we have listed them in the above table. 

It can be perceived from Table \ref{tab4} that the RCC values for $F_{\alpha \beta}$ and $K^{MS}_{\alpha \beta}$ agree well with the values extracted from the IS measurements \cite{Counts,hur}. Agreement between our calculations with experimental results for all the properties in Tables \ref{tab1} and \ref{tab4} propound about reliability of the calculated IS constants using the RCC theory. By adding nuclear deformation contributions from Ref. \cite{Allehabi} to our $F$ values, it yields $F_{\alpha}=-15569(50)$ MHz fm$^{-2}$ and $F_{\beta}=-15391(50)$ MHz fm$^{-2}$. These revised values lead to $F_{\alpha \beta} =1.0116$, which is very close to the measurement 1.01141024(86) \cite{Counts, hur}. Combining them with the measured differential frequencies of the $\alpha$ and $\beta$ transitions among the 168 - 176 isotopes \cite{Counts, hur}, the inferred $\delta \langle r^2 \rangle$ values along with uncertainties are listed in Table \ref{tab5}. We find significant differences among the previously estimated charge radii \cite{Counts, Angeli} and in this work, owing to large discrepancies in the used $F$ values.

In summary, energies and atomic clock response properties such as magnetic dipole hyperfine structure constants and quadrupole moments of the $6s~^2S_{1/2}$, $5d~^2D_{3/2}$ and $5d~^2D_{5/2}$ states of $^{171}$Yb$^+$ are determined precisely by employing relativistic coupled-cluster theory. Isotope shift constants are estimated to quite high-precision using two computationally challenging approaches in the finite-field and analytical response frameworks, which are further used to infer the nuclear charge radii of ytterbium isotopes. Considerable improvements in these values are seen due to inclusion of higher-order electron correlations through the triple excitations. We, thus, foresee a promising opportunity to probe signature of a beyond standard model vector boson by extending our method to one more step to investigate nonlinear isotope shift effects in Yb$^+$, which we defer to the future work.     

We acknowledge ANRF grant no. CRG/2023/002558 and Department of Space, Government of India for financial supports. Calculations were carried out using the ParamVikram-1000 HPC of the Physical Research Laboratory (PRL), Ahmedabad, Gujarat, India.

\end{document}